
\input harvmac

\nopagenumbers
\null\vskip-.5cm
{\hfill\ CERN-TH-6235/91}\par
{\hfill\ September 1991}
\vskip2cm
\centerline{\bf MANIFESTLY $O(d,d)$ INVARIANT APPROACH}
\centerline{\bf TO SPACE-TIME DEPENDENT STRING VACUA }
\vskip.5cm
\centerline{\bf K.A. Meissner$^*$ and G. Veneziano }
\centerline{\it Theory Division, CERN}
\vskip2cm

\centerline{\bf Abstract}

An $O(d,d)$ symmetry of the manifold of string vacua that  do not depend
  on $d$ (out of $D$)
  space-time coordinates has been recently identified. Here we write
down, for $d=D-1$,
the low energy equations of motion
and their general  solution in a manifestly $O(d,d)$-invariant
 form,  pointing
out an  amusing similarity with the
  renormalization group framework. Previously considered
cosmological and black hole  solutions are recovered
 as particular examples.

\vskip1.5in
$^*$ Permanent address: Institute of Theoretical
 Physics, ul. Ho\.za 69, 00-681
Warszawa, Poland.
\vskip.5cm
\noindent
{CERN--TH-6235/91}\par
\noindent
{September 1991}\par
\vfill
\bigskip
\eject

\def \lng{\ln\sqrt{det\ G}}
\def \gmj{{G^{-1}}}
\def \je{{\bf 1}}

\def \pmeta{\pmatrix{\ze &\je\cr \je & \ze}}
\def \ze{{\bf 0}}
\def \tra{{\rm Tr}}
\def \mdot{{\dot M}}
\def \fdot{{\dot \Phi}}

\def \fddot{{\ddot \Phi}}

\def \eps{{\epsilon}}
\def \ef{{\rm e}^\Phi}
\def \tz{{t_0}}
\def \sinh{{\rm sinh}}
\def \th{{\rm tanh}}
\def \sql{{\sqrt{\Lambda}}}
\def \ibxl{ {\int^{\infty}_{\alpha_0} {dx \over
\beta(x)} <\infty}}
 \def \ibxe{ {\int^{\infty}_{\alpha_0} {dx \over
\beta(x)} =\infty}}
\def \ibxxe{ {\int^{\infty}_{\alpha_0} {dx x \over
\beta(x)} =\infty}}
 \def \ibxla{ {\int^{\als}_{\alpha_0} {dx \over
\beta(x)} <\infty}}
 \def \ibxea{ {\int^{\als}_{\alpha_0} {dx \over
\beta(x)} =\infty}}
 \def \als{{\alpha^*}}

 {\bf 1. INTRODUCTION}\vskip1cm

 Narain's original idea
\ref\nar{
K. S. Narain, Phys. Lett. B169 (1986) 41.}, \
\ref \nswit{
K. S. Narain, M. H. Sarmadi and E. Witten, Nucl. Phys. B279 (1987) 369.} \
 of an  $O(d,d)$ symmetry acting on the manifold of static,
$d$-dimensional string compactifications, has been recently generalized
\ref\kamgv{K.A. Meissner and G. Veneziano, Symmetries of
  Cosmological Superstring Vacua, CERN-TH.6138/91,
  to appear in Phys.Lett.B.}
to the case of massless backgrounds (metric, torsion and dilaton)
 which simply
do not depend upon a particular set of  (possibly non-compact) string
coordinates $X^a (a= 1,2,...d) $.

A $\sigma$-model  argument for the validity of such a
 symmetry at all orders
in $\alpha '$ (and possibly at higher genus as well) runs
 as follows \kamgv :
Consider a conformal background (a string vacuum) of the
 kind specified above:
the associated nilpotent BRST operator $Q$ will depend  trivially (i.e.
quadratically at most) on the  $2d$ phase-space variables
$Z = (P_a , X'^a)$.   Perform now
\kamgv\ a   global, canonical   $O(d,d)$    transformation    on the
$Zs$. Since this transformation preserves commutation relations and
Wick contractions, the new BRST operator will also be nilpotent.
However, as it turns out, the change in $Z$ can be traded for a
change in the backgrounds, implying that
 also the transformed backgrounds
define a (generally inequivalent) conformal
 theory. $O(d,d)$ is thus a symmetry
of this particular class of string vacua.

Depending on the   theory at hand,  a   subgroup
of $O(d,d)$, which we shall call the |gauge
 group| $\cal {G}$, leaves the theory
unchanged.  Inequivalent vacua are thus points
 in the coset $O(d,d)/\cal {G}$.
 While
for the static, fully compact case one
 has \nar , \nswit\ the $d^2$-dimensional space
given by $\cal{G}$ $= O(d) \bigotimes O(d)$, for
the completely non-compact (but time-dependent) case  the
coset has dimensionality   $d(d-1)/2$  \kamgv  . It has been  argued
   by Sen  \ref\asen{A. Sen, $O(d)\bigotimes O(d)$ Symmetry of the
Space of
 Cosmological Solutions
in String Theory, Scale Factor Duality, and
 Two Dimensional Black Holes, Tata
Institute preprint, TIFR/TH/91-35.} that, in this latter
case, our coset is equivalent to $O(d)\bigotimes O(d)/O(d)$
and that the $O(d)
\bigotimes O(d)$ symmetry persists to all orders in $\alpha' $.
For the sake of
generality we shall base our discussion  here on the full
original $O(d,d)$ symmetry.

Discrete subgroups of $O(d,d)$ (somewhat improperly
called |duality transformations|) had been previously
discussed  both in the context of cosmological solutions
  \ref\gab{G.Veneziano, Scale Factor Duality for Classical
 and Quantum Strings, CERN-TH-6077/91.},
 \ref\tsdwa{
A. A. Tseytlin, Duality and Dilaton, Mod. Phys. Lett. 6A (1991) 1721;
 Space-Time Duality, Dilaton and
 String Cosmology, to appear
in the Proceedings of the First International A. D. Sakharov's Conference
 in Physics, Moscow, May 1991. }
and in
that of $2D$ black-holes \ref\bhs{E.Witten, On String Theory and Black
Holes,  IASSNS-HEP-91/12.},
\ref\bhb{G. Mandal, A.M. Sengupta and S.R. Wadia,
 Classical Solutions of 2-Dimensional
 String Theory, IASSNS-HEP-91/10; see also,
K. Bardakci, A. Forge and E. Rabinovici, Nucl. Phys. B344 (1990) 344;
S. Elitzur, A. Forge and E. Rabinovici, Nucl. Phys. B359 (1991) 581;
I. Bars and D. Nemeschansky, Nucl. Phys. B348 (1991) 89. },
 \ref\bhd{A. Giveon,
Target Space Duality and Stringy Black Holes, LBL-30671;
E. Smith and J. Polchinski, preprint, UTTG-07-91;
R. Dijkgraaf, H. Verlinde and E. Verlinde, preprint LBL-30747 (1991).}.
   Finally, the complete
$O(d,d)$ group has been employed  to generate, from some known
solutions, radically new candidate conformal backgrounds  both
in the black-hole \ref\sen{A. Sen, Twisted Black $p$-brane Solutions in
String Theory,  Tata
Institute preprint, TIFR/TH/91-37.} and in the cosmological \ref\gmv{M.
Gasperini, J. Maharana and G. Veneziano, From Trivial
 to Non-Trivial Conformal
String
Backgrounds via $O(d,d)$ Transformations,preprint CERN-TH-6214/91  }
 case. An
interesting aspect of $O(d,d)$, made apparent by these
 constructions, is that it
can transform gauge-equivalent backgrounds into gauge-inequivalent
 ones, as
evidentiated by the presence of a non trivial torsion field $B$ in the
transformed solutions.

In this paper we shall continue our work \kamgv\  by
 rewriting,  for $d=D-1$,  the field equations in a manifestly
$O(d,d)$-invariant form.This allows us to give
 the general solution of the (lowest order) equations
  in terms of quadratures.
Actually, equations (and solutions) take the form of renormalization-group
equations (and solutions) in real cosmic time both for
 the coupling constant of
string theory and for the matrix descibing the metric
 and torsion backgrounds.
 This analogy is used in order to obtain a general classification of the
solutions.   Finally, we shall   recover several known backgrounds as
particular cases of our general solution.
\break

{\bf 2. $O(d,d)$ SYMMETRIC EQUATIONS OF MOTION AND THEIR}

{\bf\ \ \ \ \    GENERAL
SOLUTION} \vskip1cm

In this Section we shall derive manifestly
$O(d,d)$-invariant equations of motion for the low energy effective action
of string gravity coupled to the dilaton and the antisymmetric tensor:
 \eqn\gena{S = \int d^D x\ \sqrt{-G}\ e^{-\phi} \left[\Lambda -R
        - G^{\mu\nu} \partial_{\mu}\phi\partial_{\nu}\phi -
         {1\over 12} H_{\mu\nu\rho} H^{\mu\nu\rho} \right] }
Here $\Lambda$ is the cosmological constant (proportional to D-10),
 $\phi$ is the (Fradkin-Tseytlin) dilaton
 \ref\frts{
E. S. Fradkin and A. A. Tseytlin, Phys. Lett. 158B (1985) 316;
 E. Witten, Phys. Lett. 149B (1984) 351.  }
$G_{\mu\nu}$ is the $\sigma$-model metric and $H_{\mu\nu\rho}$ is
expressed, as usual, in terms of  $B_{\mu\nu}$ via :
\eqn\hmnr{H_{\mu\nu\rho}=\partial_{\mu}B_{\nu\rho} + {\rm cyclic} }
If   $G$ and $B$ are only functions of time, invariance under
  general coordinate
 transformations and under $B_{\mu\nu} \rightarrow B_{\mu\nu} +
+\partial_{[\mu}\Lambda_{\nu ]}$ allows
always to  bring them  to   the form:
\eqn\gmun{G = \pmatrix{-1 & 0 \cr
                       0  & G(t) \cr} ,\ \ \ \ \
         B = \pmatrix{ 0 & 0 \cr
                       0  & B(t) \cr} }
Hereafter we shall use the symbols $G$ and $B$ to denote the
                $d\times d$ matrices appearing in \gmun .

 As shown in \kamgv , by introducing a $2d\times 2d$ matrix $M$
and a field $\Phi$ defined as

\eqn\mmat{M =  \pmatrix{\gmj &-\gmj B\cr B\gmj  & G-B\gmj  B\cr }   }
\eqn\bphi{\Phi = \phi - \lng.}
 the action \gena\ takes the form:
\eqn\actm{
S = \int dt\ e^{-\Phi} \left\{ \Lambda +
        (\fdot )^2 + {1\over 8}\ \tra\left[\mdot \eta
\ \mdot \eta \right]\right\}}
where $\eta$ defines an $O(d,d)$ metric in off-diagonal form i.e.
\eqn\fooe{ \eta = \pmeta}
The action \actm\ is manifestly invariant under the a global $O(d,d)$
group acting as :
   \eqn\tras{\Phi\rightarrow\Phi ,\ \ \ M\rightarrow\Omega^T M\Omega }
\eqn\omet{\Omega^T \eta \Omega = \eta}
For $d<D-1$ and backgrounds $G$  and $B$
 which are not block-diagonal, the $\sigma$-model argument given in Sect.1
 still suggests an $O(d,d)$ symmetry with the off-diagonal blocks of $M$
transforming s  by the obvious left or right multiplication with $\Omega$.

We will now obtain manifestly
$O(d,d)$-invariant equations of motion from the action \actm .
The first equation follows from reintroducing $G_{00}$
in the action and from setting to zero the corresponding variation. This
gives directly the "Zero Energy" condition
\eqn\delgzm{ (\fdot )^2 + {1\over 8}\ \tra\left[\mdot \eta
\ \mdot \eta \right]  -V = 0 }
where we allow now for a more general potential $V(\Phi)$ rather
than just a constant. If we assume that the potential does not break
the symmetry explicitly then it can depend on $M$ only through a
function of the invariants Tr$(M\eta )^p, p=1,2,...$. However, for $p$
odd these traces  vanish and for $p$ even they are equal to $2d$;
hence the potential can depend only on $\Phi$.

      The variation of the action with respect to $\Phi$
 yields:
\eqn\delfim{ (\fdot )^2 -2{\fddot }
 - {1\over 8}\ \tra\left[\mdot \eta \ \mdot \eta \right]
  +{\partial V\over \partial \Phi} -V = 0 }

The variation of the action with respect to $M$ has to be done
carefully, since $M$  is subject to several
constraints. The simplest way to proceed is to write
\eqn\mvar{\delta M = \Omega^T M \Omega - M}
where $\Omega = 1+\epsilon$ belongs to $O(d,d)$.
Expanding the variation of the action to terms linear in $\eps$
and using the fact that $\eps \eta$ is antisymmetric we get the equation:
\eqn\meqm{\partial_t (M\eta \mdot ) = \fdot (M\eta \mdot )}
which can be immediately integrated once to give:
\eqn\meqi{{\rm e}^{-\Phi} (M\eta \mdot) = {\rm const} = A.}
{}From its definition the constant matrix $A$ satisfies
\eqn\aeqns{A^T=-A,\ \ \ \ \ \ M\eta A = -A\eta M.}
It is obvious that eqs. \delgzm , \delfim\ and \meqm\
are invariant under the full $O(d,d)$ group defined
 by \tras\ and \omet .

Substituting \meqi\ into \delgzm\ we obtain the first order equation
for $\Phi$:
\eqn\liouv{ (\fdot )^2 = {\exp(2\Phi) \over 8}\ \tra (A \eta
)^2 +V(\Phi )  }
which can be solved by quadratures:
\eqn\quadr{ t= \int^{\Phi}_{\Phi _0} dy ({\exp(2y) \over 8}\ \tra (A \eta
)^2 +V(y ))^{-1/2}}
It can be easily shown that
the solution of \liouv\ automatically satisfies \delfim\ ,  provided
that $\fdot \not= 0$.
The solution  \quadr\ can be used next
to define a "dilaton time" $\tau$:

\eqn\tdefc{\tau =  \int^t_{\tz} \ef   \ dt'.}
In terms of $\tau$  the general solution of \meqi\ simply reads :
\eqn\modt{M(t) = \exp (-A\eta \tau ) M(\tz).}
  Hence the whole solution can be explicitly given in terms
of quadratures, which is a welcome surprise for  a complicated system
such as the one  described  by \gena .

To end this Section we would like to point out an amusing
similarity between equations \liouv , \tdefc\ and \modt\ and
the renormalization group (RG) equations. Introducing the
"reduced"  coupling constant  $\alpha$ by:
\eqn\dalp{\alpha = \ef}
we can   rewrite \liouv\ as a RG-equation for the running of
$\alpha$:
 \eqn\bfeq{{d\alpha \over dt} =
\beta (\alpha)}
where the $\beta$-function is given by:
\eqn\boal{\beta (\alpha) = \pm \alpha \sqrt{V(\alpha )
+{\alpha^2 \over 8}\tra (A\eta )^2}}
Once eq. \boal\ is solved, eq. \tdefc\ is  the usual
definition of the "RG-time" in terms of which moments of structure
functions evolve to leading order exponentially (generally with a
matrix-valued exponent because of operator mixing), i.e.
 precisely as in eq.
\modt . The standard RG analysis of UV or IR fixed
points can be applied to our system choosing, for definiteness,
the plus sign in
eq. \boal\ (this corresponds, physically, to choosing
 solutions which evolve
from weak coupling in the far past).
This analysis leads to the following classification scheme
for the general solution:

\item{a)} If $\beta(\alpha)$ has no other zero but
 the trivial one then we have
two possibilities:

\item{a${}_{1})$}
\eqn\alcon{\ibxl}
There is a singularity in $\Phi$ at a finite $t=t_c$. There can also be a
(coordinate or curvature) singularity if
\eqn\alconx{\ibxxe}
In this case the typical cosmology near $t_c$ is superinflationary
(Cf. \gab  ).

\item{a${}_{2})$}
  \eqn\alcone{\ibxe}
In this case no singularity can develop at any finite $t$.

\item{b)}If $\beta(\alpha)$ has a zero at some $\als >0$ then
 one has again two
possibilities:

\item{b${}_{1})$}
\eqn\aalcon{\ibxla}
No singularity develops at any finite $t$ and the coupling constant, after
growing to $\als$, goes back to zero after an infinite time.

\item{b${}_{2})$}
\eqn\alconea{\ibxea}
In this case the theory has a genuine late-time
fixed point $\alpha=\als$. At
late times a regime of exponential
 inflation ($M(t) \sim \exp (\als A \eta t)$)
sets in.

\vskip1cm
{\bf 3.  COSMOLOGICAL SOLUTIONS AND 2D BLACK HOLES AS PARTICULAR CASES}
\vskip1cm

In this Section we will solve the equations \liouv\ and \modt\
for some special potentials. The solutions we find include
time dependent cosmological solutions of \gab\ and 2D black
holes \bhs,\bhb, as well as some recently studied "boosted" versions
of both \gmv, \sen ,  but the family of solutions of \liouv\ and \modt\
is  much larger.

We start with the case of vanishing potential $V=0$ (critical
dimensions). In this case \liouv\ can be immediately solved to give
\eqn\vzal{\ef = \alpha = C/(T-t);\ \ \ \
     \ \ \ \ C=\sqrt{8/\tra (A\eta )^2} }
{}From \tdefc\ we get
\eqn\vzta{\tau = C \ln {T-t \over T-\tz}}
so that the solution for $M$ reads:
\eqn\vzsm{M(t) = \exp \left(CA\eta \ln {T-t\over T-\tz}\right)}
where we assumed that $M(\tz )=\je$.

To be more explicit let us assume some specific form for $A$:
\eqn\vzaddwa{A=\pmatrix{0&-A_d\cr
                      A_d&0\cr}}
where $A_d=diag(a_1,..,a_d)$.
In this case \vzsm\ reads:
\eqn\vzmdwa{M(t)=\pmatrix {diag\left[\left(
      {T-t\over T-\tz}\right)^{-2\alpha_1},..\right]&0 \cr
 0&diag\left[\left({T-t\over T-\tz}\right)^{2\alpha_1},
                       ..\right] \cr}}
where $\alpha_i = a_i/\sqrt{\Sigma a_i^2}$.
These solutions are exactly the ones discussed in \gab , \tsdwa (see also
 \ref\mul {M. Mueller, Nucl. Phys. B337 (1990) 37.}).
  For $\alpha _i = \alpha
\delta_{i1}$ we recover the special case of the
 completely flat "Milne" metric
\gmv, while, by inverting and boosting the
 corresponding $A$, we arrive at the
one-parameter family of non-trivial backgrounds discussed in \gmv .

We now want to discuss the solutions in presence of the potential
$V=\Lambda =$const.
The solution of \liouv\ can be immediately given:
\eqn\vlal{\ef = \alpha = C\sql /\sinh(\sql (T-t));\ \ \ \
     \ \ \ \ C=\sqrt{8/\tra (A\eta )^2} }
{}From \tdefc\ we get
\eqn\vzta{\tau = C\ln {\th (\sql(T-t)/2)\over\th (\sql (T-\tz)/2)}}
so that the solution for $M$ reads:
\eqn\vlsm{M(t) = \exp \left(CA\eta\ln{\th (\sql (T-t)/2)\over
\th (\sql (T-\tz)/2)}\right)}
where we again assumed that $M(\tz )=\je$.

For $d=1$ we take
\eqn\vladdwa{A=\pmatrix{0&-a\cr
                      a&0\cr} \ \ \ \ \ a>0 }
and assume that $M(-\infty)=\je$. Then
\eqn\vlmdwa{M(t)=\pmatrix {\th^{-2} (\sql(T-t)/2) &0 \cr
               0& \th^2 (\sql(T-t)/2)\cr}}
\eqn\vlefdw{\ef = {2\sql\over a\sinh(\sql (T-t))}}
After the identification \kamgv :
\eqn\idnbh{t\leftrightarrow ix,\ T=0,\ \Lambda  = -4}
we recover the $k=9/4$ black hole solution of \bhs , \bhb . Acting
 on the solution
with $\eta$ we recover the "dual" 2D black hole of ref.\bhd .

For $d>1$ we can use the same $A$ (with zeroes in the additional
 entries) and
reproduce the $k \not= 9/4$ black holes of \bhs , while,
 by boosting any of these
solutions one recovers the "twisted" black holes of \sen .

We wish to conclude by mentioning one  amusing solution that exists
for $d=9$ (usual uncompactified superstring). As $A$ we take
\eqn\ddzae{A = \pmatrix{0&diag(-a_1,..,-a_9)\cr
                 diag(a_1,..,a_9)&0\cr}}
Then $M$ is equal to
\eqn\ddzm{M=\pmatrix{diag\left(\th^{-2\alpha_1}
             (\sql (T-t)/2),..\right)&0\cr
    0&diag\left(\th^{2\alpha_1} (\sql (T-t)/2),..\right)\cr}}
 where $\alpha_i=a_i /\sqrt{\Sigma a_i^2}$.
The scalar curvature and dilaton are given by:
\eqn\ramus{R=-{\Lambda\over \cosh^2(\sql (T-t))/2)}\left[\sum \alpha_i
  -{(\Sigma \alpha_i -1)^2\over 4 \sinh^2(\sql (T-t)/2))}\right]}
\eqn\famus{{\rm e}^{\phi} = {(\th(\sql (T-t)/2))^{\Sigma \alpha_i -1}
\over {a\ \cosh^2(\sql (T-t)/2))}}.}
Hence we have a singularity for $t\rightarrow T$ unless
\eqn\sali{\sum \alpha_i =1.}
Assuming that all $\vert \alpha_i \vert$ are equal,
the only solution for $\alpha_i$ is (-1/3,-1/3,-1/3,+1/3,..,+1/3)
so that three dimensions expand and six contract!
\vskip1in
We are grateful to D. Amati, M. Gasperini and E. Rabinovici
 for discussions.

\listrefs
\end